\documentclass{aa}  

\usepackage{graphicx}
\usepackage{txfonts}
\usepackage{siunitx}
\usepackage[outercaption]{sidecap}   
\usepackage{natbib,twoopt}
\usepackage{xcolor}
\usepackage[breaklinks=true]{hyperref} %% to avoid \citeads line fills
\bibpunct{(}{)}{;}{a}{}{,}             %% natbib format for A&A and ApJ
\makeatletter
  \newcommandtwoopt{\citeads}[3][][]{\href{http://adsabs.harvard.edu/abs/#3}%
    {\def\hyper@linkstart##1##2{}%
     \let\hyper@linkend\@empty\citealp[#1][#2]{#3}}}
  \newcommandtwoopt{\citepads}[3][][]{\href{http://adsabs.harvard.edu/abs/#3}%
    {\def\hyper@linkstart##1##2{}%
     \let\hyper@linkend\@empty\citep[#1][#2]{#3}}}
  \newcommandtwoopt{\citetads}[3][][]{\href{http://adsabs.harvard.edu/abs/#3}%
    {\def\hyper@linkstart##1##2{}%
     \let\hyper@linkend\@empty\citet[#1][#2]{#3}}}
  \newcommandtwoopt{\citeyearads}[3][][]%
    {\href{http://adsabs.harvard.edu/abs/#3}
    {\def\hyper@linkstart##1##2{}%
     \let\hyper@linkend\@empty\citeyear[#1][#2]{#3}}}
\makeatother

\begin{document}

   \title{Rapid Blue- and Red-shifted Excursions in H$\alpha$ line profiles
synthesized from realistic 3D MHD simulations}

   %\subtitle{I. Overviewing the $\kappa$-mechanism}

   \author{S. Danilovic \inst{1}
          \and
          J. P. Bj{\o}rgen
          \and
          J. Leenaarts \inst{1}
          \and
          M. Rempel \inst{2}
          %\fnmsep\thanks{Just to show the usage
          %of the elements in the author field}
          }

   \institute{Institute for Solar Physics, Dept. of Astronomy, Stockholm University, Albanova University Center, 10691 Stockholm, Sweden\\
              \email{sdani@astro.su.se}
         \and High Altitude Observatory, National Center for Atmospheric Research, 80307, Boulder, CO, USA \\
             }

   \date{}

% \abstract{}{}{}{}{} 
% 5 {} token are mandatory
 
  \abstract
  {Rapid blue- and red-shifted events (RBEs/RREs) may have an important role in mass-loading and heating the solar corona, but their nature and origin are still debatable.}
   {We aim to model these features to learn more about their properties, formation and origin.}
   {A realistic three-dimensional (3D) magneto-hydrodynamic (MHD) model of a solar plage region is created. Synthetic H$\alpha$ spectra are generated and the spectral signatures of these features are identified. The magnetic field lines associated with these events are traced and the underlying dynamic is studied.}
   {The model reproduces well  many properties of RBEs and RREs, such as spatial distribution, lateral movement, length and lifetimes. Synthetic H$\alpha$ line profiles, similarly to observed ones, show strong blue- or red-shift and asymmetries. These line profiles are caused by the vertical component of velocity with magnitudes larger than $30-40$~km/s that appear mostly in the height range of $2-4$~Mm. By tracing magnetic field lines, we show that the vertical velocity that causes the appearance of RBE/RREs to appear is always associated with the component of velocity perpendicular to the magnetic field line.}
   {The study confirms the hypothesis that RBEs and RREs are signs of Alfv{\'e}nic waves with, in some cases, a significant contribution from slow magneto-acoustic mode.}


   \maketitle
%
%-------------------------------------------------------------------

\section{Introduction}

The solar chromosphere is shaped from long, dense fibrils \citep{2007ASPC..368...27R} that connect large magnetic features organized at mesogranular to supergranular scales. These magnetic features are the centers of most chromospheric dynamics  \citep{2012SSRv..169..181T}. Radially extending from those are dark stable features called mottles, visible in cores of chromospheric spectral lines. Moving to the wings of chromospheric lines, features named rapid blue- and red-shifted events (RBEs/RREs) become dominant. Unlike mottles, these features are very dynamic, short-lived, long and thin, but also visible near the edges of the magnetic network. Although it is generally believed that they have an important role in mass-loading and heating the solar corona, they are still a puzzle. 

The RBEs were first detected in the blue wing of the Ca~II~$854.2$~nm line as sudden broadening in $\lambda$t-dataslices \citep{2008ApJ...679L.167L}. Their lifetimes, spatial extent and location, as well as sudden disappearance, suggested that they might be the on-disk counterparts of then newly discovered off-limb spicule dynamics \citep{2007PASJ...59S.655D}. Subsequent papers based on observations with higher spatial and temporal resolution further strengthened this hypothesis \citep{2009ApJ...705..272R, 2012ApJ...752..108S, 2013ApJ...764..164S}. The spectral imaging data taken in both Ca~II~$854.2$~nm and H$\alpha$ lines revealed that the RBEs vary strongly between these lines. The former shows significantly fewer and shorter features appearing closer to the magnetic field concentrations and generally before the H$\alpha$ RBEs. Also, the corresponding Doppler shifts differ, being of order $20-50$~km/s in the H$\alpha$ and  $15-20$~km/s in the Ca~II~$854.2$~nm line. The differences were explained by the chromospheric opacity in the lines, with H$\alpha$ sampling higher layers and naturally faster upflows.  

The properties of these features also differ depending on the detection method used and the temporal cadence of the data. The distribution of the RBEs lifetimes ranges from $5$ to $60$~s, with a slow cadence data shifting it to the higher values. This may be due to the high recurrence times around magnetic elements which gives apparently higher lifetimes at a lower cadence. Furthermore, the overall behavior of RBEs changed with temporal resolution. The first studies claimed that RBEs moved away from magnetic concentration with Doppler shifts and widths increasing from the footpoint to the top. The higher cadence data, however, revealed much more RBEs that display erratic behavior. While there were cases that had a distinct rise phase, there were also features that showed a complicated mixture of longitudinal variations. There were many cases that seemingly came to a halt or even retreated or just appeared suddenly along the full length within a few seconds. Some RBEs moved sideways over a few hundred kilometers, while others hardly got displaced. It also became clear that RBEs appear in groups with large and dark jet-like features appearing first, followed by numerous shorter, thinner jets \citep{2013ApJ...767...17Y,2019Sci...366..890S}. 
\begin{figure*}
   \centering
      \includegraphics[width=\linewidth,trim= 0cm 0cm 0cm 1.8cm,clip=true]{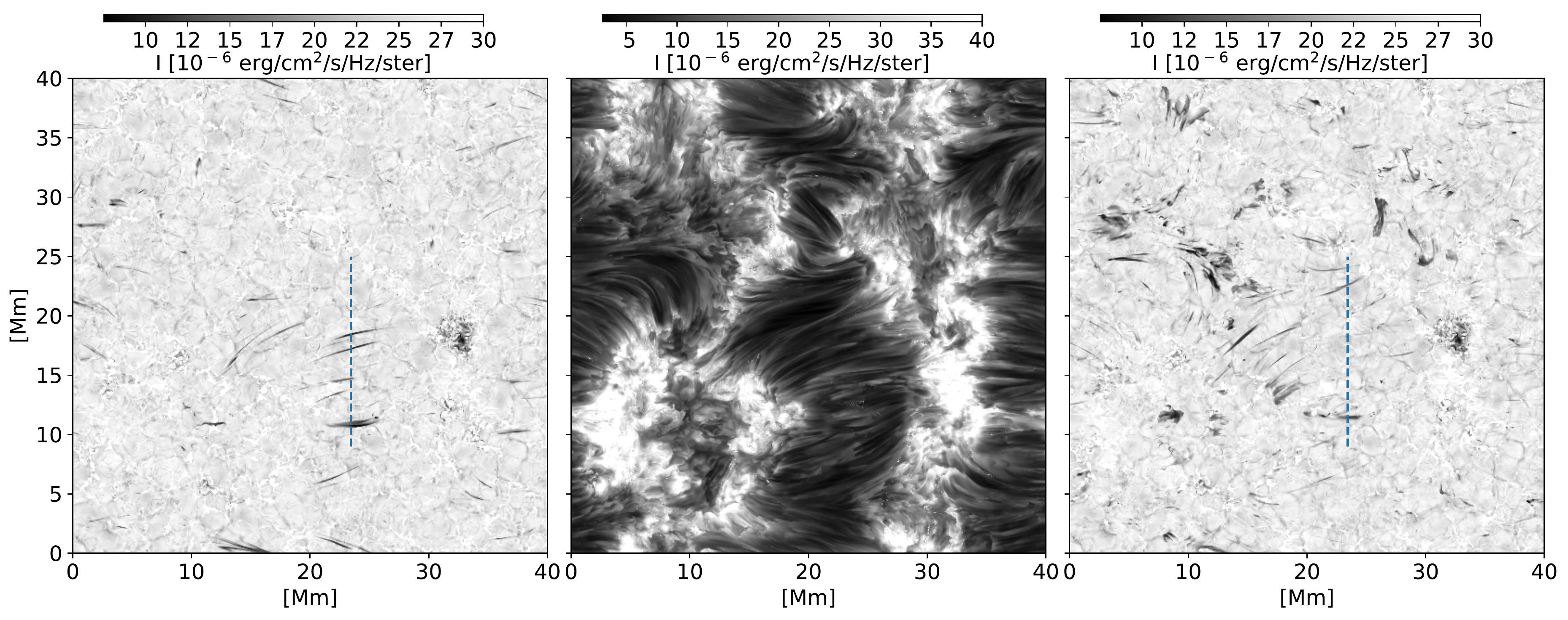}
   \caption{Synthetic H$\alpha$ images at different wavelength positions: left - blue wing  at $v_{dopp} = - 36$~km~s$^{-1}$, middle - nominal line center and right - red wing $v_{dopp} = 36$~km~s$^{-1}$. Blue vertical line marks the position of the cut shown in Fig.~\ref{cut}. The movie is available in the supplementary material.\href{https://dubshen.astro.su.se/~sdani/rbes/fig1.mp4}{[Movie]} }
        \label{overview}%
\end{figure*}

The RREs are similar absorption features but in the red wings of the Ca~II~$854.2$~nm and H$\alpha$ lines \citep{2013ApJ...769...44S, 2015ApJ...802...26K, 2016A&A...589A...3S, 2019A&A...631L...5B}. RREs are found over the whole solar disk and are located in the same regions as RBEs. Lengths, widths, lifetimes, and average Doppler velocity of RREs are similar to RBEs. RREs have similar occurrence rates as RBEs with the total occurrence rate increasing towards the limb. Most RREs and RBEs are observed in isolation, but many examples of parallel and touching RRE/RBE pairs are also found to be part of the same spicule. \cite{2015ApJ...802...26K} detect more RREs and  \cite{2013ApJ...769...44S} more RBEs. The latter study also shows that the RRE/RBE detection count ratio increases from the disk center to the limb from $0.52$ to $0.74$. The higher number of RBEs and the decreased imbalance towards the limb is interpreted as an indication that field-aligned upflows have a significant contribution to the net Dopplershift of the structure.

Transverse displacements of RREs and RBEs are also similar. There are examples of transitions from RRE to RBE and vice versa that sometimes appear to propagate along the structure. In some cases, even oscillatory behavior can be found with periods averaging $90$~s and amplitudes of $200$~km. All these characteristics are in line with a hypothesis that these are an indication of Alfv{\'e}nic waves associated with the swaying motion. Namely, the movement of the 'flux tube' is detected in both image planes as transverse motion and along the line-of-sight (LOS) as a sudden blue- and red-shifts in the wings of spectral lines. The formation of such phenomena would also arise naturally in case of magnetic reconnection, which was indicated by observations that show RBEs generally appearing when new, mixed or unipolar fields are detected in close proximity to network fields \citep{2013ApJ...767...17Y, 2015ApJ...799..219D, 2019Sci...366..890S}. Although this scenario explains all peculiarities of RBEs and RREs, alternative hypotheses must also be mentioned. One alternative explanation is that the observed transverse motions are not waves, but just transverse displacement of the flux tube as a whole from its initial position as it happens when the interaction timescale between the granule and the flux tube is comparable to or greater than the chromospheric cutoff period \citep{1999ApJ...519..899H}. Another alternative explanation is that the sudden shifts in the spectral lines can be an indication of sheet-like structures \citep{2012ApJ...755L..11J, 2014ApJ...785..109L}. 

There is a large number of numerical experiments that try to explain the formation and origin of spicules \citep[e.g.][]{1969PASJ...21..128U, 1982ApJ...257..345H,1982SoPh...75...99S, 2010A&A...519A...8M,2012ApJ...744L...5J,2017ApJ...848...38I, 2017Sci...356.1269M}. However, only two studies tried to explicitly model RBEs and RREs \citep{2015ApJ...802...26K,2017NatSR...743147S}. These models fail to reproduce many properties of RBEs and RREs. Still, they indicate that RBEs/RREs may be signatures of Alfv{\'e}n waves generated by torsional driving in the photosphere. The present manuscript describes a three-dimensional (3D) model that reproduces spatial distribution, lateral movement, length and lifetimes of RBEs and RREs, as well as  strongly blue- or red-shift asymmetric H$\alpha$ line profiles. We start with identifying the RBE/RRE signatures in synthetic spectra generated from the numerical models. We then compare their properties with observations and further study their formation.

\section{Model atmospheres and forward synthesis}
The MURaM code \citep{2005A&A...429..335V, 2017ApJ...834...10R} is used to generate the numerical 3D models of the solar atmosphere. The model is built in phases starting from the nonmagnetic convection simulation to which a uniform bipolar field of $200$~G is added. After the field configuration evolved, the potential field extrapolations are used to extend the computational domain and include the upper atmosphere. The final extent of the simulation domain is $40 \times 40 \times 22$~Mm, spanning vertically from $  -8$~Mm to $14$~Mm above the
average $\tau_{500}= 1$ height. At the bottom boundary, a horizontal field at roughly equipartition field strengths is allowed to 
emerge into the domain \citep{2014ApJ...789..132R}. At the upper boundary, the field is set to be vertical. Also, the upper boundary
is open to outflows but closed to inflows. The horizontal boundaries are periodic. In the last phases, the model is first
run at a lower resolution and then again at the double resolution until a relaxed state is achieved. The grid spacing of the simulation used in this paper is $39$~km and $21$~km in the horizontal and vertical directions, respectively. The model contains only the most basic physics to be called ‘comprehensive’: 3D grey LTE radiative
transfer, a tabulated LTE equation of state, Spitzer heat conduction, and optically thin radiative losses in the corona based on
CHIANTI \citep{2012ApJ...744...99L}. The numerical diffusive and resistive terms in this run are set so that an effective magnetic Prandtl number P$_{\rm m} > 1$. This makes viscous heating to be the largest contributor. The maximum
Alfv{\'e}n velocity is chosen as max$(2$c$_{s}, 3|$v$_{ \rm max}|)$ where v$_{ \rm max}$ is the maximum velocity and c$_{\rm s}$ the speed of sound \citep{2017ApJ...834...10R}. This gives the largest unaffected Alfv{\'e}n velocity of $900$~km/s \citep{2021arXiv210614055C}. The snapshots are outputted every $~2.5$~s and every fourth snapshot is used as input for the subsequent radiative transfer computations. 

For these computations, we used Multi3D (\citet{2009ASPC..415...87L}) with the model atmosphere that we cropped in two ways. The vertical span of the input model is limited to $360$ points which include the range of temperatures relevant for the formation of the H$\alpha$ line. In the horizontal direction, we use every second grid point to make computations faster. Tests show that this does not affect the appearance of synthetic features. H$\alpha$ was calculated using a 3-level plus continuum H model atom created by \citet{2019A&A...631A..33B} to increase numerical stability in Multi3D. The synthetic spectra are generated for $29$ snapshots in total.

% [right, bottom, right, top]
\begin{figure}

   \centering

    \includegraphics[width=0.8\linewidth,trim= 0cm 0cm 0cm 0cm,clip=true]{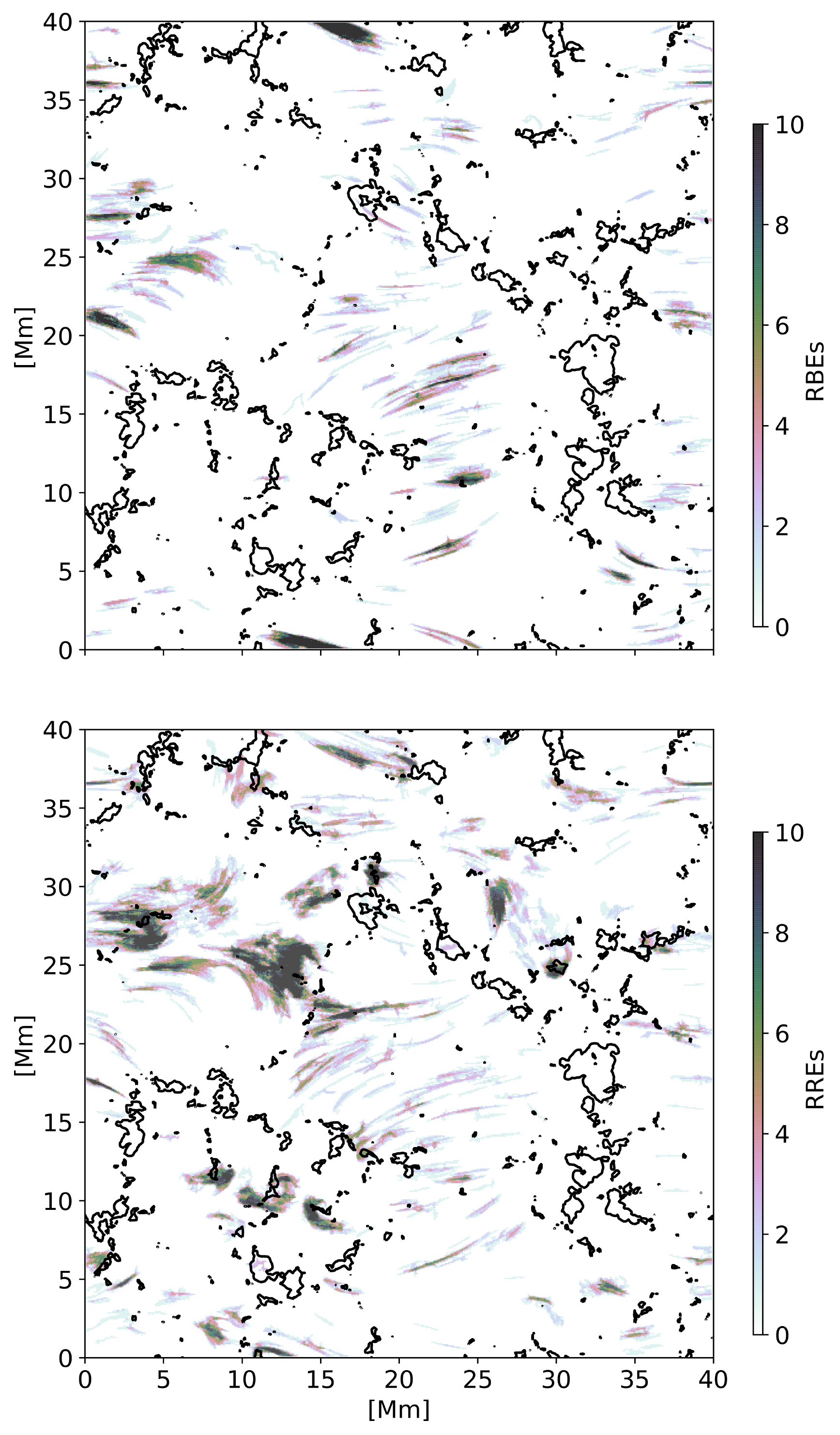}  
   \caption{Automatically identified synthetic RBEs and RREs. The panels show summed masks for RBEs and RREs, respectively. The black contours outline the unsigned photospheric vertical field of $1000$~G. }
        \label{mask}%
\end{figure}
% angle=90 [bottom ,right, top, left]
\begin{figure}
  \includegraphics[angle=90,width=0.52\linewidth,trim= 0.8cm 1.5cm 1.7cm 2.1cm,clip=true]{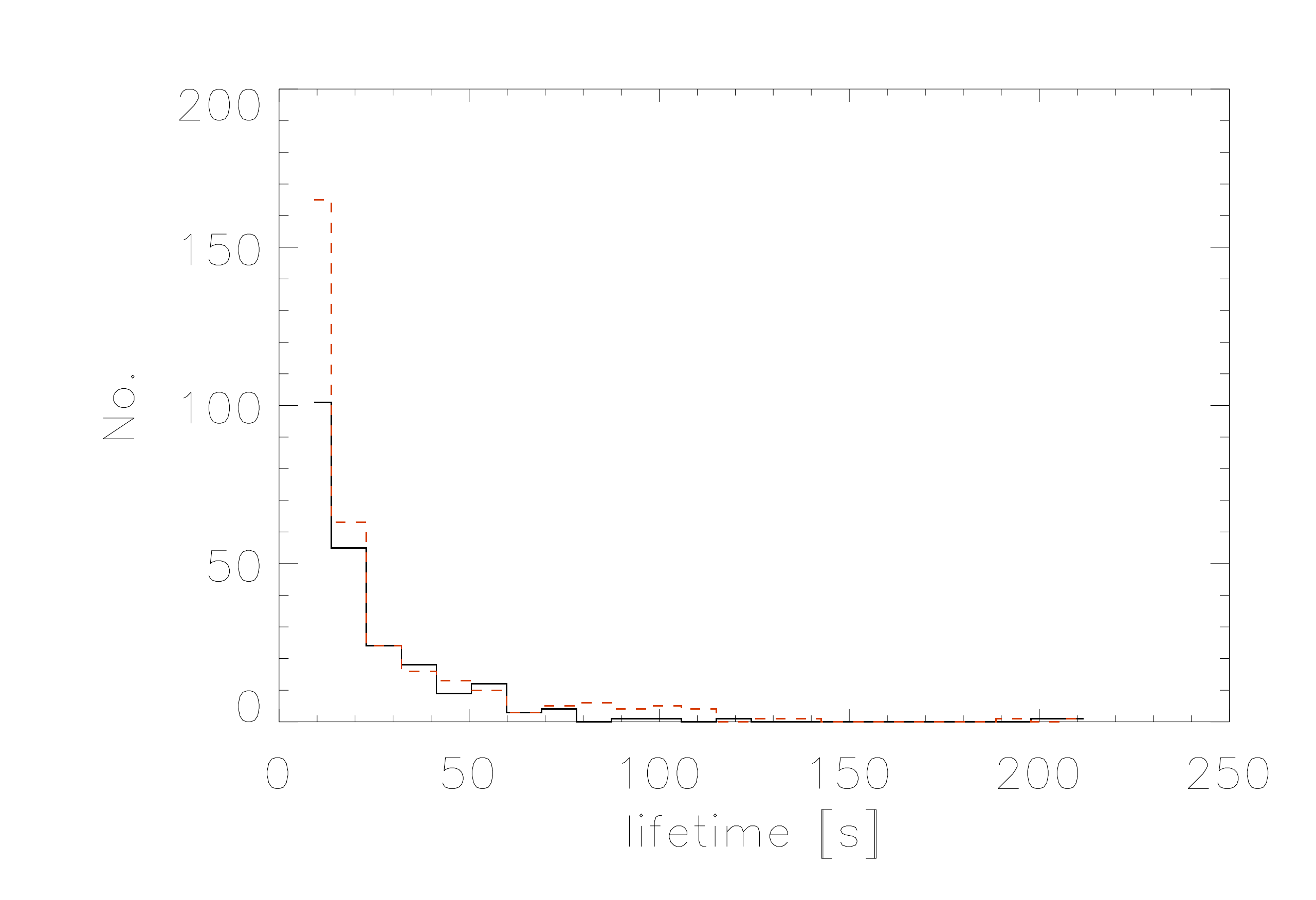} 
      \includegraphics[angle=90,width=0.45\linewidth,trim= 0.8cm 1.5cm 1.5cm 5.3cm,clip=true]{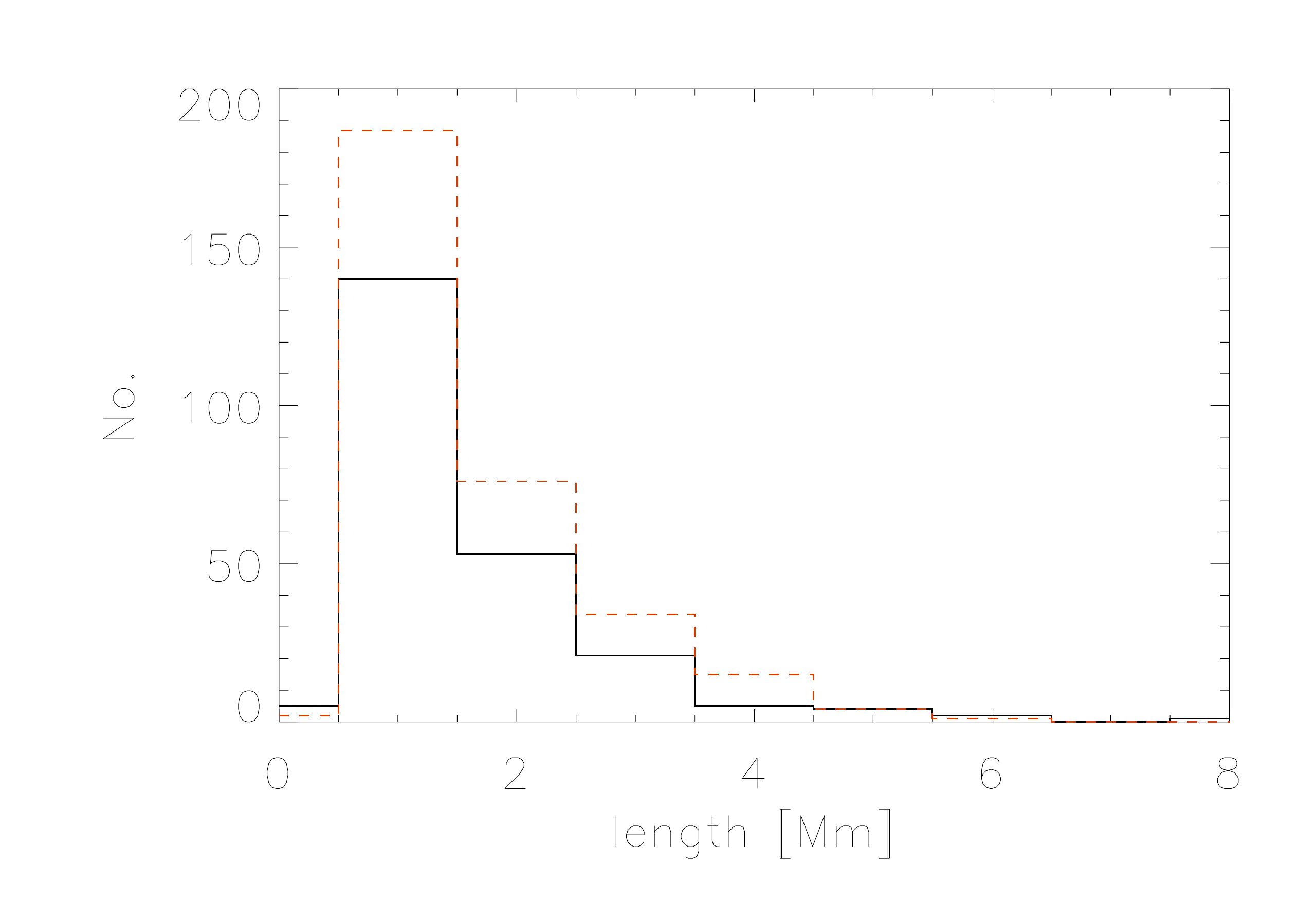}
   \caption{Lifetime (left) and length (right) of tracked synthetic RBE (black) and RREs (red).}
        \label{prop}%
\end{figure}

\section{Results}
\subsection{Spatial distribution of synthetic RBEs and RREs}

Figure~\ref{overview} shows synthetic H$\alpha$ images generated from one of the snapshots. The magnetic field configuration is visible from the distribution of long fibrils in the H$\alpha$ core shown in the middle panel. Panels on the sides show synthetic RBEs and RREs. The wavelength positions correspond to Doppler shifts of $\pm 36$~km/s, similar to observations. Thin elongated dark features are visible in both H$\alpha$, wings, resembling their observed counterparts. Besides thin features, the red-wing image shows slightly more spread-out structures especially close to the 'slope' generated by the inclined magnetic field around $[7,25]$~Mm. The dark circular structure at $[33,20]$~Mm visible in both wings is a pore. 

Inspection of the movie corresponding to Fig.~\ref{overview} reveals different kinds of features. In both RBE and RRE images, there are examples that show apparent motion or increase in length away and towards magnetic network. There are also many examples that appear and disappear in situ, seemingly halfway between the magnetic footpoints. Many features show transverse motion. All this agrees with findings based on observations.   

To study the statistics of the synthetic RBEs and RREs, we apply simple filtering on the H$\alpha$ wing images, exposing all regions where the intensity was below $22.3 \times 10^{-6}$~erg/cm$^{2}$/s/Hz/ster and aspect ratio larger than 15. These numbers are somewhat arbitrarily chosen. The final masks summed over $29$ snapshots are visible in Fig.~\ref{mask}. The figure shows the occurrence and feature position with respect to magnetic elements, which are outlined with black contours. The chosen criteria allow the inclusion of fairly short features as the one at $[3,30]$~Mm in the top panel. It also allows many extended structures visible in the H$\alpha$ red wing images to be counted as RREs, as the bottom panel Fig.~\ref{mask} shows. Although magnetic elements are distributed all over the simulation domain, maps nevertheless show areas with reduced activity at $[25,2.5]$, $[28,15]$ and $[35,30]$~Mm. In other regions, RREs and RBEs can be found in seemingly equal numbers. The exception is visible at the base of the 'slope' where RREs prevail. 

\subsection{Properties of synthetic RBEs and RREs}
We cluster masked features that overlap from one snapshot to the other and count them as one feature. This gives the total number of RBEs and RREs to be 231 and 322, respectively. The distribution of their lifetimes is shown in Fig.~\ref{prop}. Many features traced this way appear and disappear in one snapshot, giving them lifetimes of less than $10$~s. Most of the extra RREs is placed into this bin. The distribution exponentially decreases with the lifetime. Only around $5$~\% of both RRE and RBEs seems to be visible for at least a minute.     

For each case, we track the evolution and include their most extended length in the distribution shown on the right of Fig.~\ref{prop}. We see that most features are shorter than $3$~Mm in size, although some cases extend to over $5$~Mm. There seems to be no preference for the extra short-lived RREs. They seem to be distributed proportionally to all length bins. 

\subsection{Formation of synthetic RBEs and RREs}

\begin{figure}
   \centering
   \includegraphics[width=\linewidth,trim= 1.75cm 0.5cm 1.85cm 0cm,clip=true]{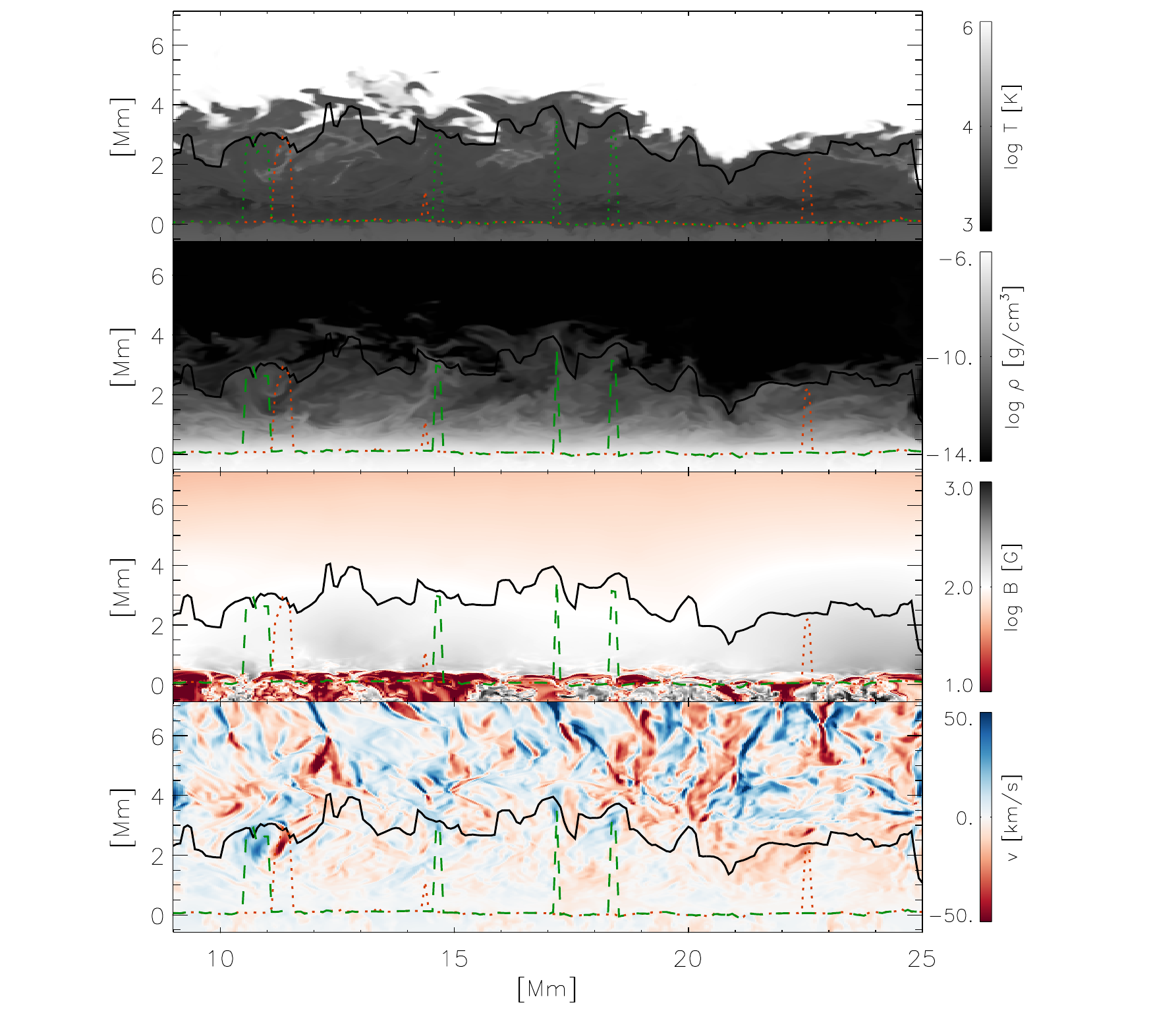}
   \caption{Vertical cut along the blue vertical lines shown in Fig.~\ref{overview}. Vertical cuts in temperature, density, magnetic field strength and vertical velocity are shown from top to bottom. The black line in all panels shows the $\tau =1$ at the nominal position of the H$\alpha$ line center. Green and red contours mark $\tau =1$ levels at wavelength positions that correspond to Doppler shifts of $v_{ \rm dopp} = - 36$~km~s$^{-1}$ and $v_{\rm dopp} = 36$~km~s$^{-1}$ respectively. }
        \label{cut}%
\end{figure}

Figure~\ref{cut} shows a vertical cut over several RBE and RRE at the time instant shown in Fig.~\ref{overview}. Overploted are $\tau =1$ contours for the same wavelength positions as in Fig.~\ref{overview}. Corrugated black lines thus show that the H$\alpha$ line core is formed in the range $2-4$~Mm, while jumps in green and red contours indicate the positions of the crossed RBEs and RREs respectively. Most of them, the four RBEs and two RRE, are formed close to the H$\alpha$ line core.They are formed at similar heights, which results in similar intensities visible in Fig.~\ref{overview}. The bottom panel of Fig.~\ref{cut} shows that at each position where RBE/RREs is formed, there is a strong upflow/downflow in the vertical velocity map. The field strength in those regions is still higher than $100$~G, as the third panel in Fig.~\ref{cut} shows.
% [right, bottom, right, top]

\begin{figure}[h!]
   \centering
   \includegraphics[width=\linewidth,trim= 0.65cm 0.45cm 0.6cm 0.4cm,clip=true]{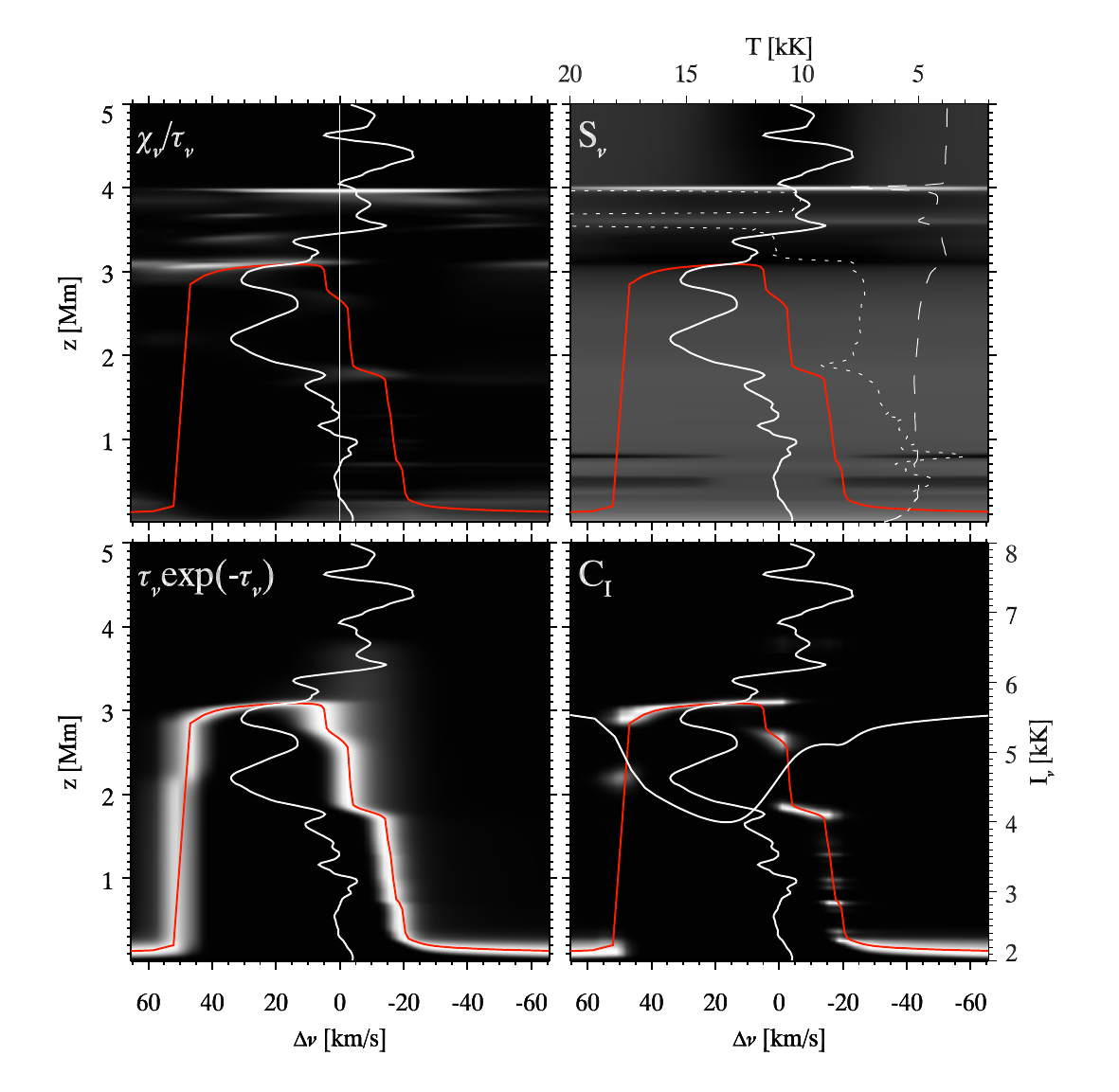}
   \caption{The H$\alpha$ four-panel diagram for the RBE at $x=10.7$~Mm in Fig.~\ref{cut}. The plotted quantities, as functions of frequency and height, are indicated in the upper left corner of each subplot and plotted in a grayscale. The vertical component of velocity as a function of height is shown in all panels (solid white line), with upwards velocity being positive. The red line marks the $\tau =1$ position as a function of frequency and height. Planck and total source function are plotted in the upper right panel with dotted and dashed lines, respectively. The line profile is shown with solid white line in lower right panel.
   }
        \label{RBE_prof}%
\end{figure}
\begin{figure}[h!]
   \centering
   \includegraphics[width=\linewidth,trim= 0.65cm 0.45cm 0.6cm 0.4cm,clip=true]{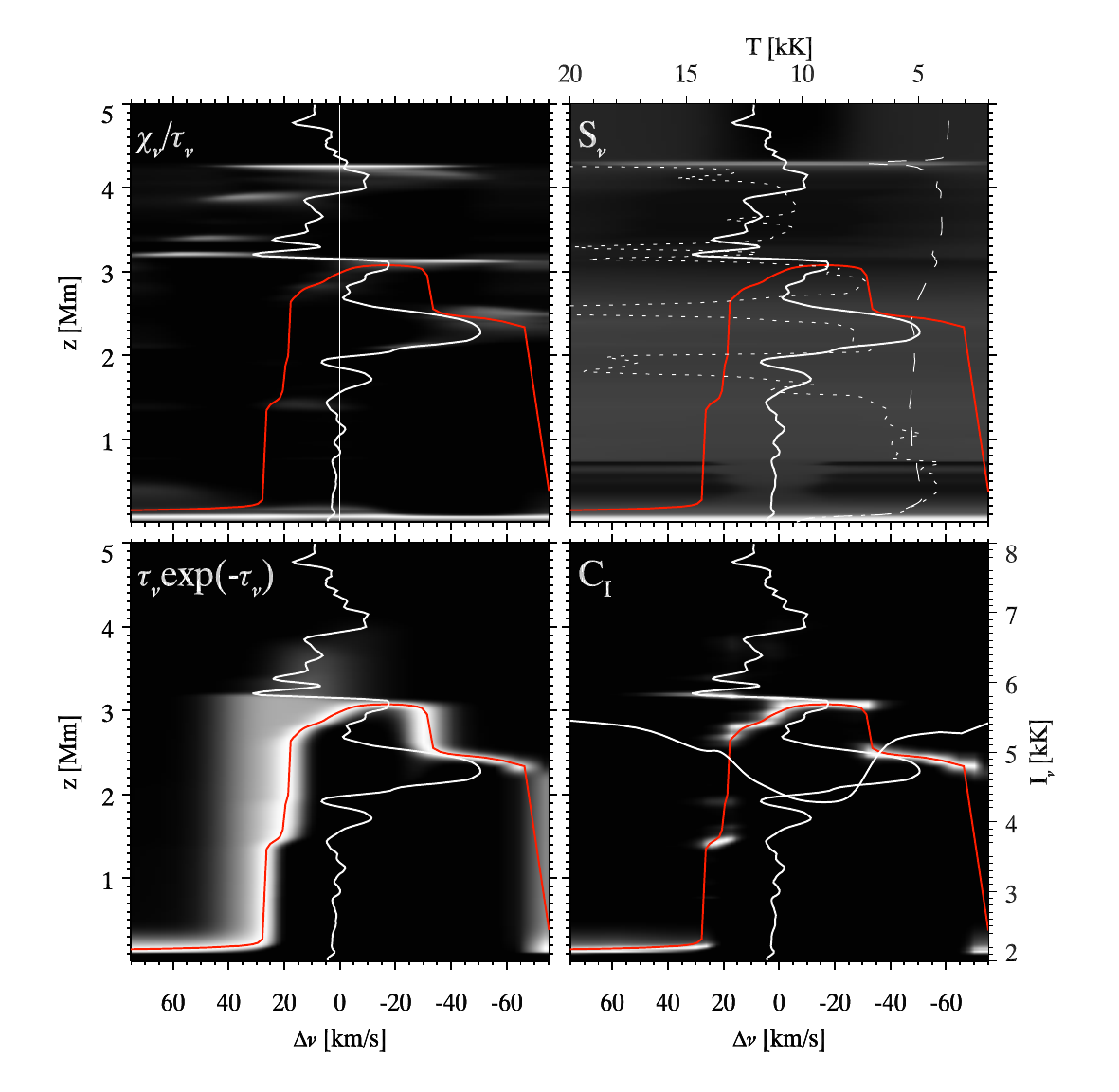}
   \caption{The H$\alpha$ four-panel diagram for RRE at $x=11.25$~Mm in Fig.~\ref{cut}. The outline is the same as in Fig.~\ref{RBE_prof}.
   }
        \label{RRE_prof}%
\end{figure} 
Examples at $x=10.7$~Mm and $x=11.25$~Mm in Fig.~\ref{cut} are analyzed in more detail using four-panel schematic H$\alpha$ line formation diagrams \citep{1997ApJ...481..500C, 2010ApJ...709.1362L} shown in Figs.~\ref{RBE_prof} and \ref{RRE_prof}. An example of an RBE is shown in Fig.~\ref{RBE_prof}. The emergent profile in the lower right panel is strongly blue-shifted with a small reversal at the red wing. Overplotted in the same panel is the line-of-sight (LOS) velocity profile displaying a shift in the same direction. The predominant upflows reaching $40$~km/s are spread in the height range $2-3$~Mm. Because of these upflows, the location with high $\chi_\nu / \tau_\nu$  (upper left panel) is strongly shifted at z$=3$~Mm, exactly where the optical depth reaches unity at those frequencies (lower left panel). The source function is completely decoupled from the Planck function and stays fairly flat with two small local maxima higher up, at $3.5$ and $4$~Mm. The opacity has a shape characteristic for the H$\alpha$ line with sharp, sudden shifts of $\tau = 1$ in both wings but strongly shifted to the positive frequencies. As a result, the contribution function jumps to the height of $3$~Mm and stays there over a wide range of frequencies. A small reversal in the red wing is formed due to a temperature and density increase that are also visible in Fig.~\ref{cut}. The formation mechanism of the red wing reversal is the same as for spectra signatures of the so-called Ellerman bombs \citep{2017A&A...601A.122D}.

The four-panel diagram created for the pixel at x$=11.25$~Mm in Fig.~\ref{cut} is presented in Fig.~\ref{RRE_prof}. This pixel experiences a much more complex opacity distribution than in the case of Fig.~\ref{RBE_prof}. Here, we have numerous bright regions in the upper left panel at a variety of altitudes and frequencies. Two major effects on the contribution function come from high  $\chi_\nu / \tau_\nu$ located at z$=3$ and z$=2.3-2.5$~Mm. The latter is the location where a strong downflow reaches $50$~km/s. At the former location, there is a switch from an upflow to a downflow, both slightly weaker reaching $20$~km/s. The source function shows a local maximum in both locations. As a result, line wings and core intensities are all experiencing contributions from both heights. The blue-shifted material also coincides with the location where the optical depth reaches unity at that frequency, so the blue wing gets extended, resulting in a very asymmetric line profile. 
% [right, bottom, right, top]
\begin{figure}
 \centering
\includegraphics[width=\linewidth,trim= 0.2cm 0cm 8.5cm 0cm,clip=true]{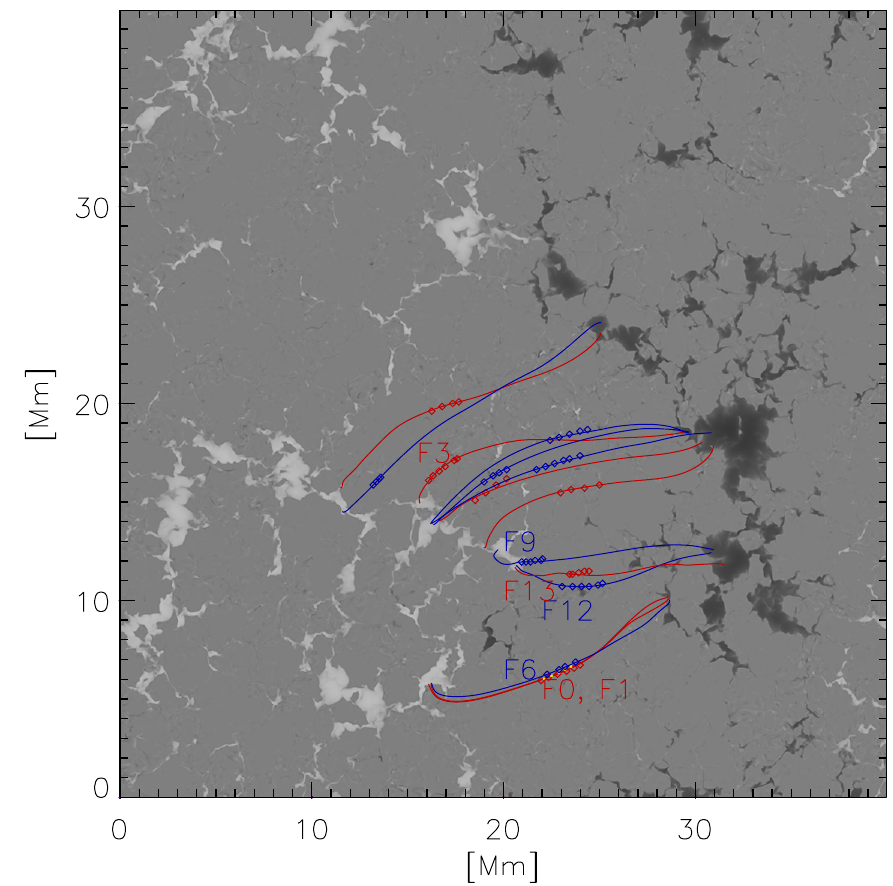} 
  \caption{The overview of the RBE/RRE sample that is chosen for magnetic field line tracing. For each of the $14$ examples, a traced magnetic field line is plotted over the vertical component of the photospheric magnetic field. The diamonds mark the position of the starting seed used for line tracing: red marks RREs and blue RBEs. Yellow stars mark the position of the F1. }
\label{line_trace}
\end{figure}
% angle=90 [bottom ,right, top, left]
\begin{figure}
 \centering
\includegraphics[angle=90,width=\linewidth,trim= 1.5cm 0.3cm 5cm 0.3cm,clip=true]{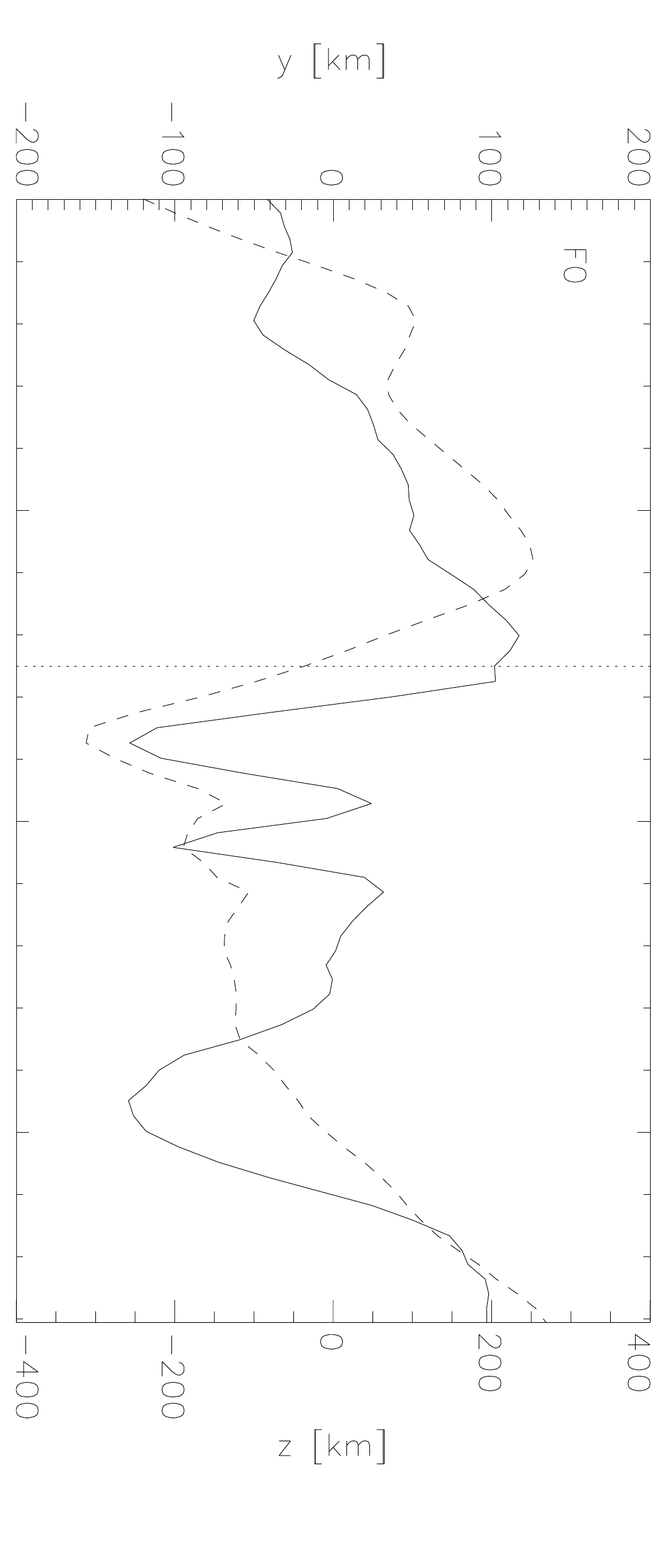} 
\includegraphics[angle=90,width=\linewidth,trim= 1.5cm 0.3cm 5cm 0.3cm,clip=true]{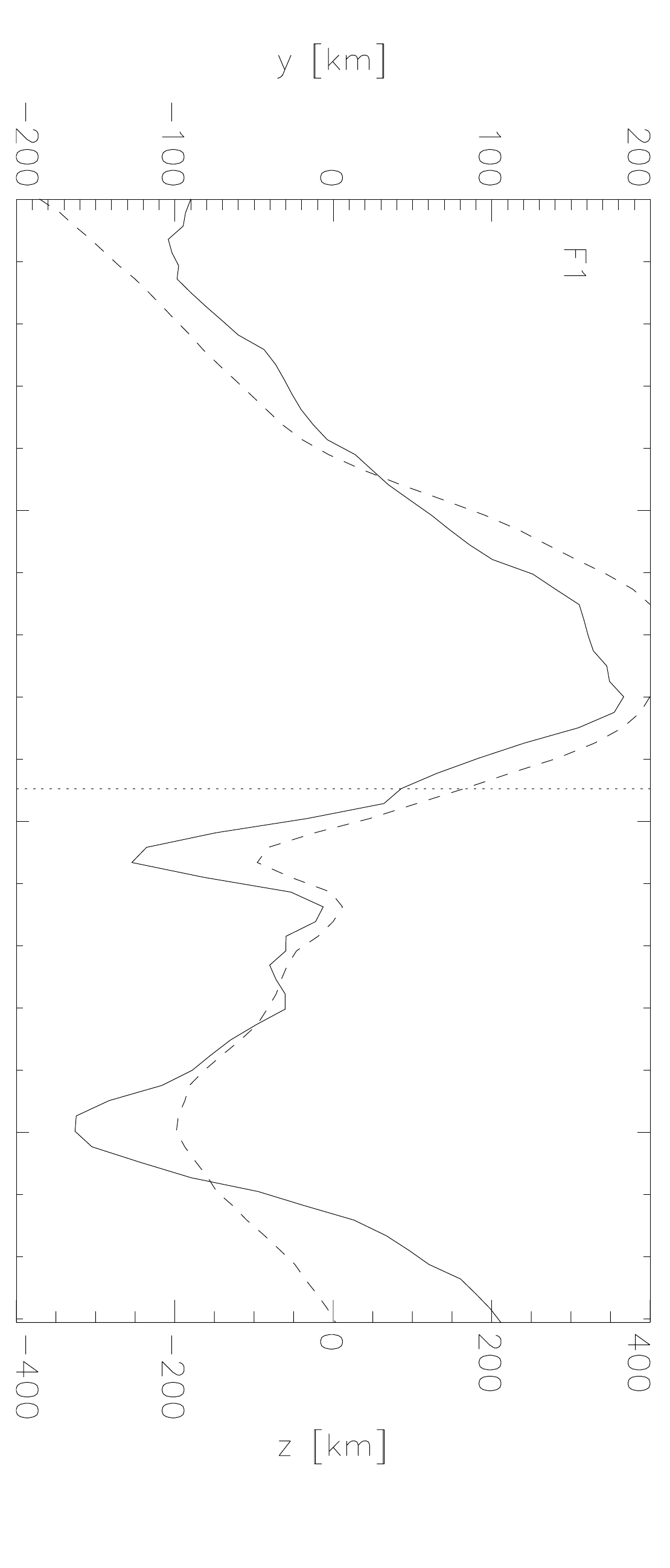} 
\includegraphics[angle=90,width=\linewidth,trim= 1.5cm 0.3cm 5cm 0.3cm,clip=true]{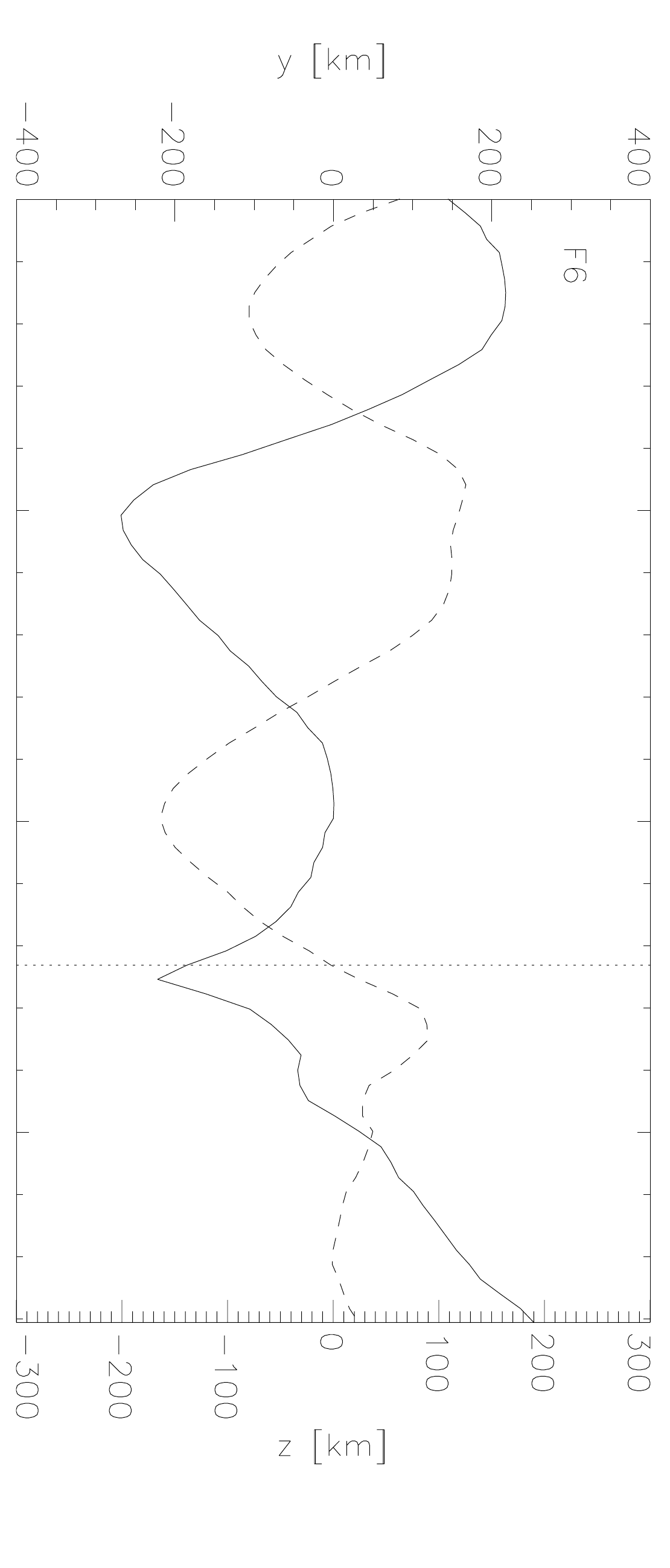} 
\includegraphics[angle=90,width=\linewidth,trim= 0.1cm 0.3cm 5cm 0.3cm,clip=true]{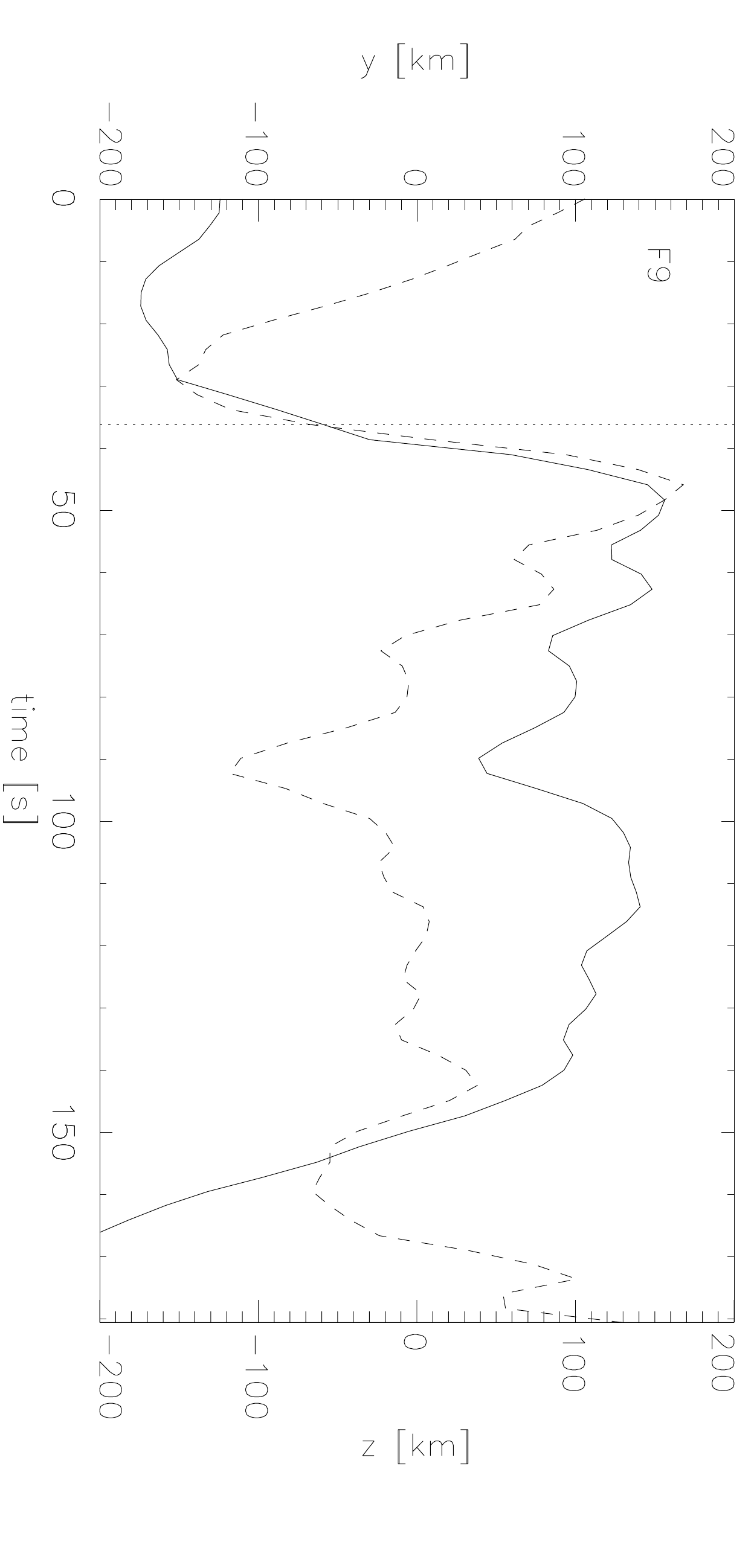} 
  \caption{The horizontal (y, solid line) and vertical (z, dashed line) displacement at constant x for $4$ examples of RBE/RRE. The vertical line marks the time when the seed is 'planted'. The cases F0 and F1 are RREs and F6 and F9 RBEs. The labels are the same as in Fig.~\ref{line_trace}.}
\label{yz_shift}
\end{figure}

\subsection{Origin of synthetic RBEs and RREs}

We further choose $14$ examples around the center of the simulation domain to study what causes their appearance. An equal number of RBEs and RREs are chosen. They appear at different time instances distributed over $9$ different snapshots. Figure~\ref{line_trace} gives an overview of positions of one traced field line per each chosen feature. We trace the magnetic field lines as curvatures in 3D space \citep{2015ApJ...802..136L}. The time cadence between snapshots used for the field tracing is around $2.5$~s. For each of the examples, we pick several seeds along the feature, visible either in blue or red synthetic H$\alpha$ wing, to advect as corks forward and backward in time. For each seed, a formation height at that specific wavelength position is taken as the z-coordinate in the 3D space. The individual cases are labeled with F$0-13$ as shown in Fig.~\ref{line_trace}. Cases F12 and F13 are the RBE and RRE visible in Fig.~\ref{cut} and further discussed in four-panel diagrams. The yellow stars in Fig.~\ref{line_trace} mark the position of the RRE associated with F1, which appears close to F0 when F0 disappears. The line tracing indicates that these RREs belong to the same event since their origin can be traced to overlapping field lines\footnote{The movie is available as supplementary material.}. 
% angle=90 [bottom ,right, top, left]
\begin{figure*}[h!]
 \centering
 \includegraphics[angle=90,width=\linewidth,trim= 1.6cm 1.7cm 9cm 0.2cm,clip=true]{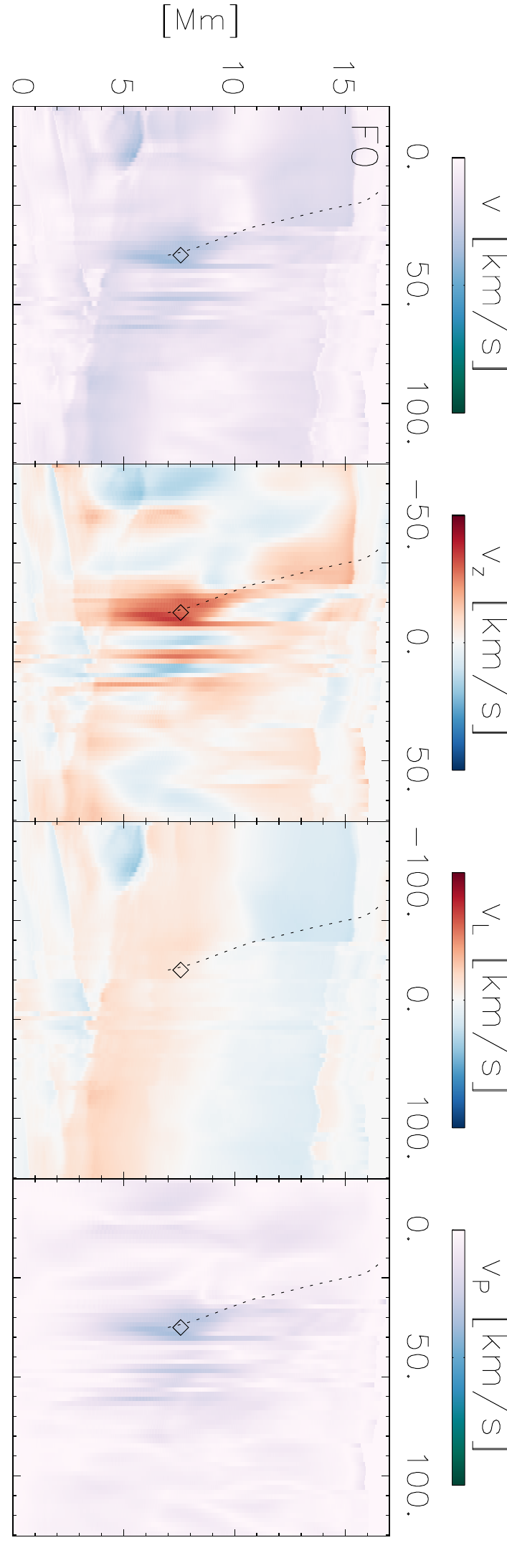} 
  \includegraphics[angle=90,width=\linewidth,trim= 1.6cm 1.7cm 10.5cm 0.2cm,clip=true]{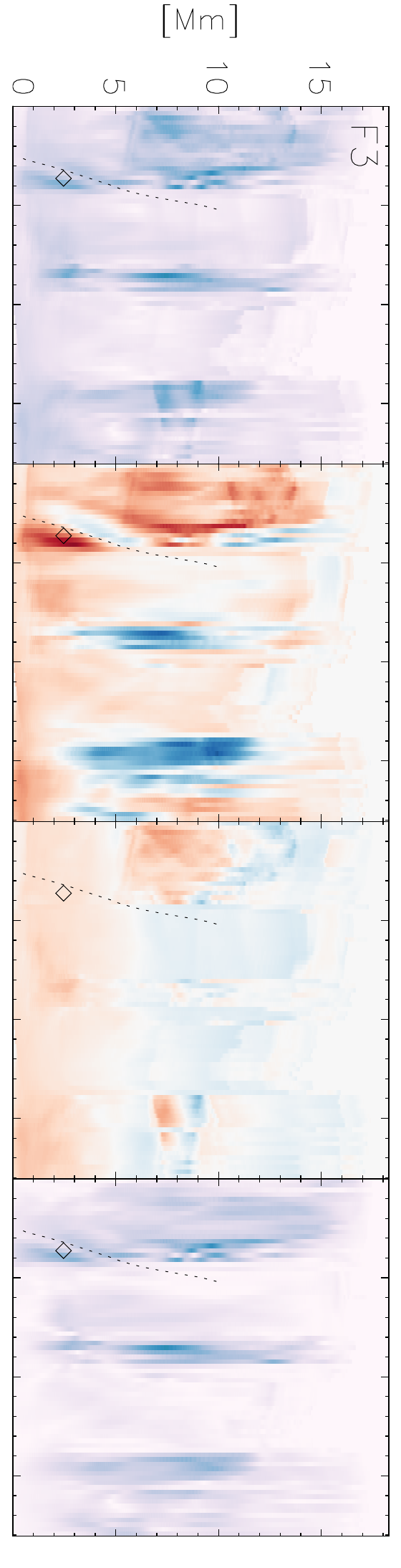} 
\includegraphics[angle=90,width=\linewidth,trim= 1.6cm 1.7cm 10.5cm 0.2cm,clip=true]{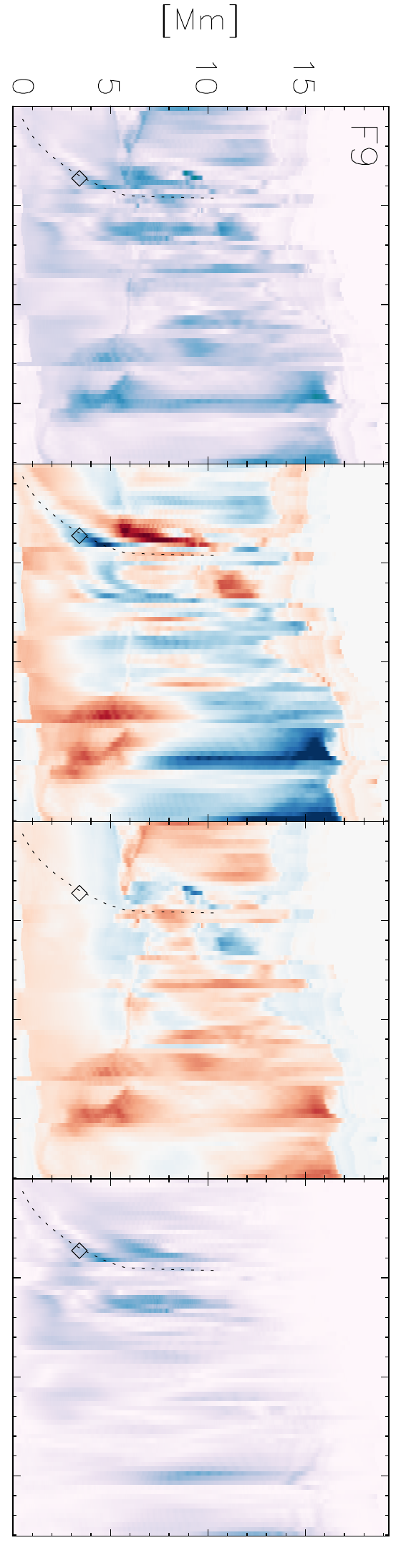} 
 \includegraphics[angle=90,width=\linewidth,trim= 0.5cm 1.7cm 10.5cm 0.2cm,clip=true]{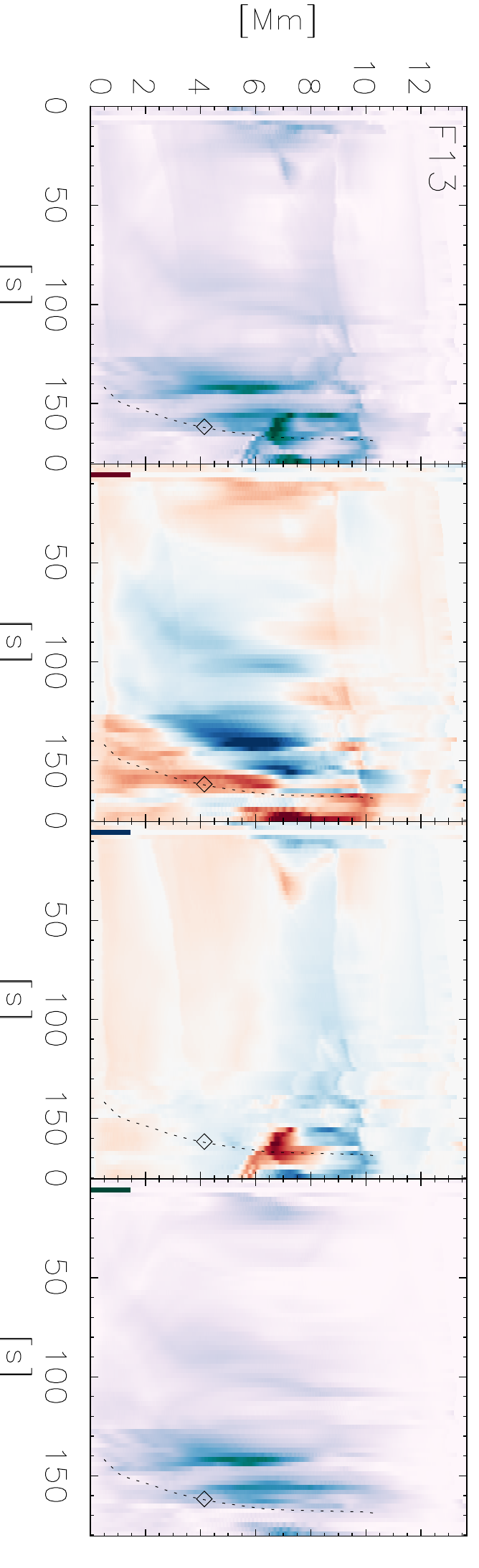} 
  \caption{The evolution along a field line traced for four cases labeled in the top left corner. From left to right: total velocity v, vertical velocity v$_{\rm z}$, longitudinal v$_{\rm L}$ and perpendicular v$_{\rm P}$ velocity. Diamonds mark the location of starting seeds. The dotted lines follow a signal propagating with the Alfv{\'e}n speed along the loop. }
\label{F_time}
\end{figure*}

To analyze the displacement of each chosen example, we place the 'slit' at the x coordinate of the starting seed and then follow the y and z coordinates of the traced field line in time at that fixed x position. Figure~\ref{yz_shift} shows resulting horizontal and vertical displacements for four cases after a linear fit is subtracted. The shifts in two directions get in the phase when the RBE/RRE appears and sometimes show oscillatory behavior over a short time. For example, F0 shows a sinusoidal wave pattern in y coordinate that lasts for $\sim 40$~s, while in the case of F9, the same lasts $\sim 100$~s. In most cases, we see more erratic behavior like for F6. The four cases shown in Fig.~\ref{yz_shift} illustrate that the range of displacements stays within $200$~km with amplitudes of a few $10$~km during the oscillatory period.    
% [right, bottom, right, top]
\begin{figure*}
   \centering
      \includegraphics[width=0.9\linewidth,trim= 0cm 1.35cm 0cm 0cm,clip=true]{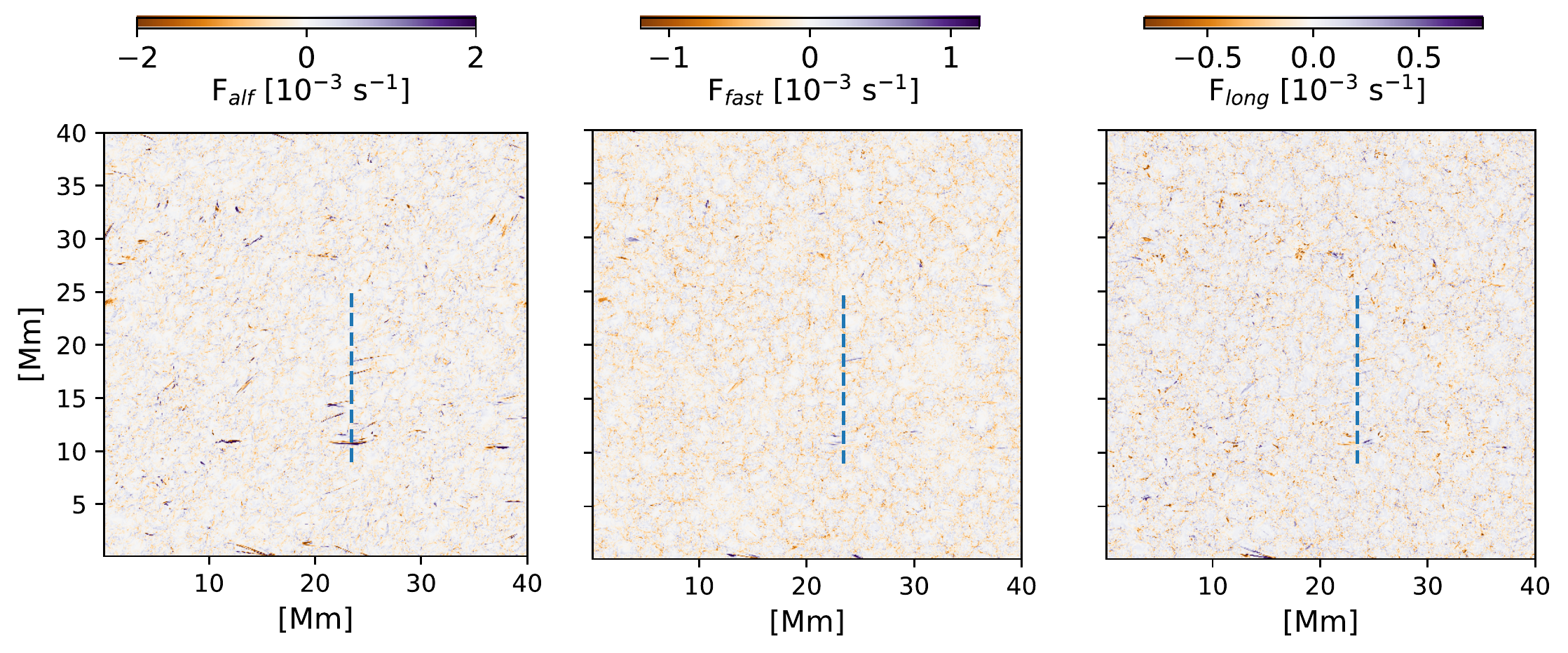}
     \includegraphics[width=0.9\linewidth,trim= 0cm 1.35cm 0cm 1.8cm,clip=true]{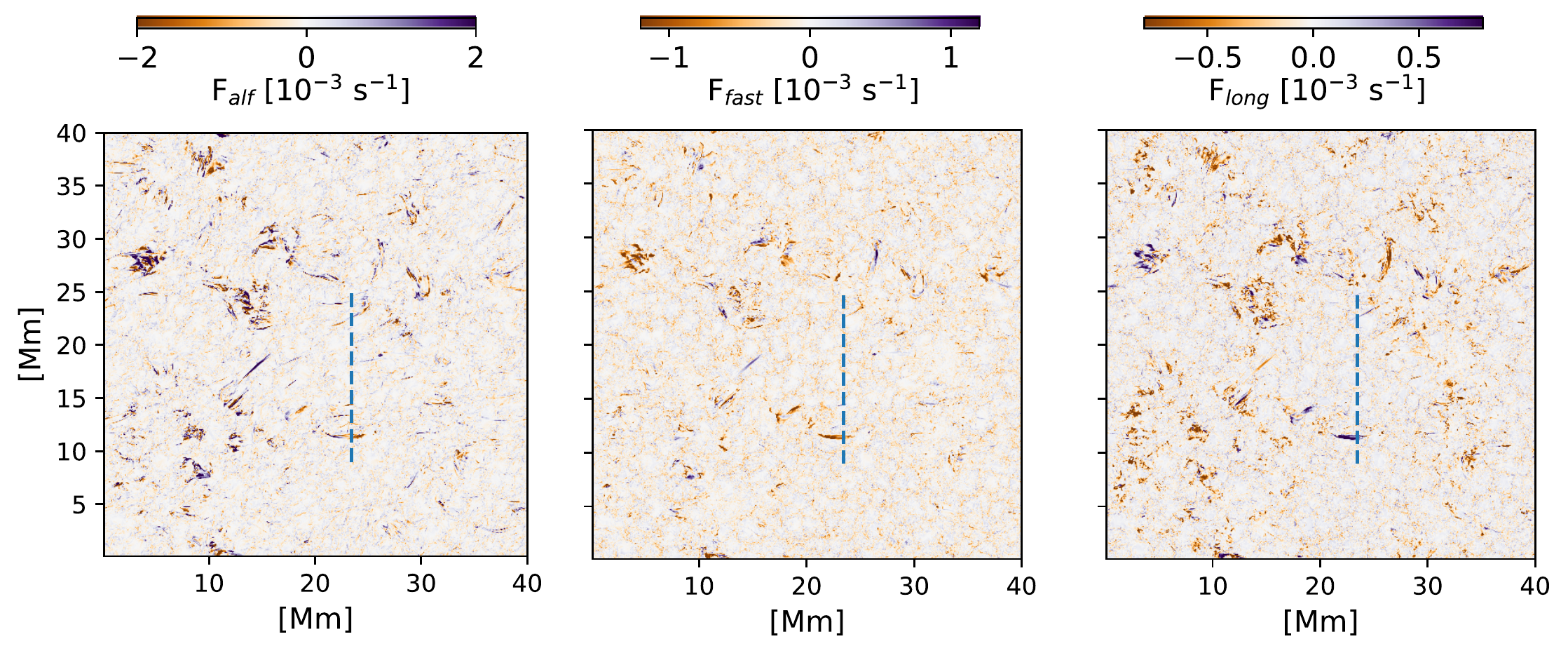}
      \includegraphics[width=0.9\linewidth,trim= 0cm 8.4cm 0cm 0cm,clip=true]{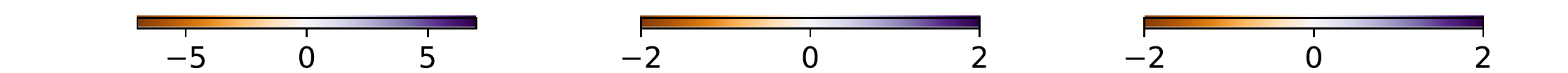}
     \includegraphics[width=0.9\linewidth,trim= 0cm 0cm 0cm 1.8cm,clip=true]{fwave_ha.eps}

   \caption{Proxies of different wave modes: Alfv{\'e}n (left column), fast (middle column) and slow mode waves convolved with the contribution function at different wavelength positions: top row - blue wing  at v$_{d \rm opp} = - 36$~km~s$^{-1}$, middle row -  red wing v$_{ \rm dopp} = 36$~km~s$^{-1}$ and bottom row - nominal line center. Blue vertical line marks the position of the cut shown in Fig.~\ref{cut}. }
        \label{fwave}%
\end{figure*}
The field tracing reveal two common characteristics for all cases. This is illustrated in Fig.~\ref{F_time} on four RBE/RRE examples. The figure shows velocities along a traced field line as a function of time. Diamonds mark the location of starting seeds and dotted lines follow signals propagating with the Alfv{\'e}n speed along the loop. The first common characteristic for all cases is that we detect a strong flow at one of the footpoints which perturbs the field line. This is visible in the vertical component of velocity displayed in the second column. For cases F3, F9 and F13, alternating upflow and downflow are visible close to the first footpoint. For case F0, the same happens at the second footpoint. The flow propagates along loops with Alfv{\'e}n speed of $ \sim 300$~km/s for all inspected cases. The second common characteristic is that the vertical velocity that causes RBE/RREs to appear in H$\alpha$ line wings is always associated with the component of velocity perpendicular to the magnetic field line. This is visible in the third and fourth column, where we see high-velocity pattern only in v$_{\rm P}$ and nothing in v$_{\rm L}$ at the time instant in proximity to the position of the starting seed. Only in the case F13, strong flow is visible along the field line v$_{\rm L}$, although not at the position of this specific seed. 

We can also use instantaneous proxies for three wave modes slow $f_{long}$ and fast $f_{fast}$ magneto-acoustic and Alfv{\'e}n waves $f_{alf}$  \citep{2018A&A...618A..87K, 2017MNRAS.466..413C}:
\begin{align}
f_{alf}= \hat{e_{\parallel}}\cdot \bigtriangledown \times \mathbf{v} \\
f_{fast}= \bigtriangledown \cdot \left ( \mathbf{v} -  \hat{e_{\parallel}}v_{\parallel} \right ) \\
f_{long}= \hat{e_{\parallel}} \cdot \bigtriangledown \left ( \mathbf{v} \cdot \hat{e_{\parallel}} \right )  
\end{align}
where $\mathbf{v}$ is the velocity vector and $\hat{e_{\parallel}}$ the field-aligned unit vector. To visualize where the wave power for each mode is located with respect to the formation of RBEs and RREs, each quantity is convoluted with a contribution function at the corresponding wavelength. The resulting maps are shown in Fig.~\ref{fwave}.

There is a significant difference between maps corresponding to the line center and the line wings. The latter are largely contaminated with the contribution from the photosphere. Only in regions where RBEs and RREs occur the contribution function shifts to higher layers, as illustrated in the previous section, and the maps show elongated features. If we compare Fig.~\ref{fwave} with Fig.~\ref{overview}, we can identify many features in the intensity. Most of them are visible only in $f_{alf}$. In some cases, weak signatures can be also traced in $f_{fast}$ maps. A few cases can also be identified in all three wave proxies. Cases F12 and F13 belong in the last group of features. There also seems to be a trend that $f_{long}$ and $f_{fast}$ signatures are located closer to the footpoints and $f_{alf}$ proxy extends further along a feature \citep[see also][]{2022arXiv220803744D}.

\section{Discussion and conclusions}

H$\alpha$ line profiles generated from 3D model of solar plage show RBEs and RREs signatures. Their properties: spatial distribution, length and lifetimes are very similar to the properties of their observed counterparts. The synthetic features appear close to the magnetic network or at the apex of the loops. Similarly to the observations, their behavior can be both consistent or erratic. In the first case, they move away or closer to the network. In the second case, they appear and disappear far from any magnetic concentration. They show lateral motion of a few $10 - 100$~ km. This agrees with observations \citep{2012ApJ...752..108S,2013ApJ...764..164S}. We identify cases where features exhibit lateral motion by being offset along the whole length, as reported in observations \citep{2015ApJ...802...26K}. An example of this is the feature labeled F0 and F1. 

Synthetic H$\alpha$ line profiles associated with these features are either completely blue- or red-shifted or they show asymmetric, extended wing. These line profiles are caused by the vertical component of velocity with magnitudes larger than $30-40$~km/s. In our model, these velocities appear mostly in the height range of $2-4$~Mm, at the line core formation height or slightly below that. By tracing magnetic field lines, we show that the vertical velocity that causes RBE/RREs to appear in H$\alpha$ line wings is always associated with the component of velocity perpendicular to the magnetic field line. Further analysis shows that features mainly outline the proxy location of  Alfv{\'e}n waves and to a lesser degree of fast magneto-acoustic waves. In some cases, proxies of all three wave modes can be found coinciding with the RBE/RREs formation. As most events were associated to Alfven waves, the hypothesis that RBEs and RREs are signs of Alfv{\'e}nic waves \citep{2013ApJ...764..164S} is confirmed.

Tracing the magnetic field lines also demonstrates that strong flows at one of the footpoints perturbs field lines, as a result of which RBE/RREs are formed. This explains why RBEs appear in groups \citep{2013ApJ...767...17Y, 2019Sci...366..890S}. Strong jets would generate perturbations in all adjacent field lines which would result in numerous shorter, thinner features like the ones modeled in this study. Our model, however, does not show many cases with strong jets. Likewise, the model generates more RREs than RBEs. This is inconsistent with observations. We find several possible reasons for this. Our detection method included many features that might not be counted as RRE if found in the observations. Among these are the ones that resemble dynamic fibrils. These features, when observed, show $5$ minute oscillations in the region with a more inclined field \citep{2007ApJ...655..624D}. Our synthetic H$\alpha$ dataset covers less than that so there is a possibility that we caught most of them in their descending phase. This will be the subject of our next study.

The resolution of our model can also be a reason for generating RBE/RRE ratio different than the one observed. Namely, the higher resolution would produce more turbulent flows, resulting in footpoint motions that could lead to more frequent magnetic reconnection. As the reconnection would happen near footpoints, the upflows would prevail in the height ranges where RBE/RREs form and thus generate more RBEs than RREs. Also, flux recycling or small-scale emergence generated at higher resolution would result in fast lateral movement and would have the same effect. Although most of the flux emerging on granular scales gets pulled back \citep{2010ApJ...723L.149D}, we do not exclude the scenario where these features are generated by the reconnection of emerging with the pre-existing field, which our simulations do not model. Finally, the velocity magnitude increases with resolution, hence more events will result in sufficient line shifts. The second possible reason for the mismatch of simulations and observations could be in not including the ion-neutral interaction effects i.e. ambipolar diffusion \citep{2017Sci...356.1269M}. With ambipolar diffusion included, the plasma is not fully frozen in, so the jets may be more vertical, not following the field lines. This would lead to a larger vertical velocity component and the identification of more jets. 

Apart from the unresolved cause of the mismatch between simulated and observed RBE/RREs ratio, a few more questions remain for the following studies. One of which is what determines the width of these features. In models like \cite{2017NatSR...743147S} or in a cartoon by \cite{2013ApJ...769...44S}, RBEs and RREs are generated at the side edges of magnetic flux tubes. Based on this interpretation, observations \citep{2017NatSR...743147S, 2021arXiv211214486S} would suggest that these flux tubes are resolved and sometimes as an arcsec wide. \cite{2021JGRA..12629097S} further argue that if the waveguide is not spatially resolved the opposing Doppler shifts will result in non-thermal broadening in optically thin lines while the chromospheric observables would sample only 'the outermost shells of the structure'. We find no distinct structures as such in our model, although some synthetic RBE/RREs pairs can be traced to the same source. The model shows that while H$\alpha$ line core outlines density ridges \citep{2012ApJ...749..136L, 2022arXiv220803744D}, the formation of the line wings is more complex.  

Finally, we did not discuss the photospheric source of these features in the model. Movies of traced field lines show braiding of field lines close to a footpoint. This suggest the formation of photospheric vortex flow. The vortex signatures are present in all areas of the simulation domain where magnetic field is stronger. They are most prominent in the temperature maps sampled at upper photosphere, as shown by \cite{2012A&A...541A..68M}. Vorticity caused by torsional motions in the photosphere can excite torsional Alfv{\'e}n pulses that propagate along the magnetic field lines to the upper layers in the form of torsional Alfv{\'e}n waves \citep{2013ApJ...776L...4S,2019NatCo..10.3504L, 2021A&A...649A.121B}. All this is agreement with the results presented in this study.

\begin{acknowledgements}
This project has received funding from Swedish Research Council (2021-05613), Swedish National Space Agency (2021-00116) and the Knut and Alice Wallenberg Foundation. This material is based upon work supported by the National Center for Atmospheric Research, which is a major facility sponsored by the National Science Foundation under Cooperative Agreement No. 1852977. This research data leading to the results obtained has been supported by SOLARNET project that has received funding from the European Union’s Horizon 2020 research and innovation programme under grant agreement no 824135. The calculations were performed on resources provided by the Swedish National Infrastructure for Computing (SNIC) at the National Supercomputer Centre (NSC) at Linköping University and the PDC Centre for High Performance Computing (PDC-HPC) at
the Royal Institute of Technology in Stockholm. 
\end{acknowledgements}

\appendix{

\section{Supplementary material}
Besides the movie version of Fig.~\ref{overview}, we share two additional movies that show magnetic field lines for cases F9, F0 and F1. The field lines of F0 and F1 are plotted together, as shown in Fig.~\ref{fig:F01_movie} so that initial overlapping of magnetic field lines is visible.
\begin{figure*}
\centering
\includegraphics[width=\linewidth,trim= 0cm 0cm 0cm 0cm,clip=true]{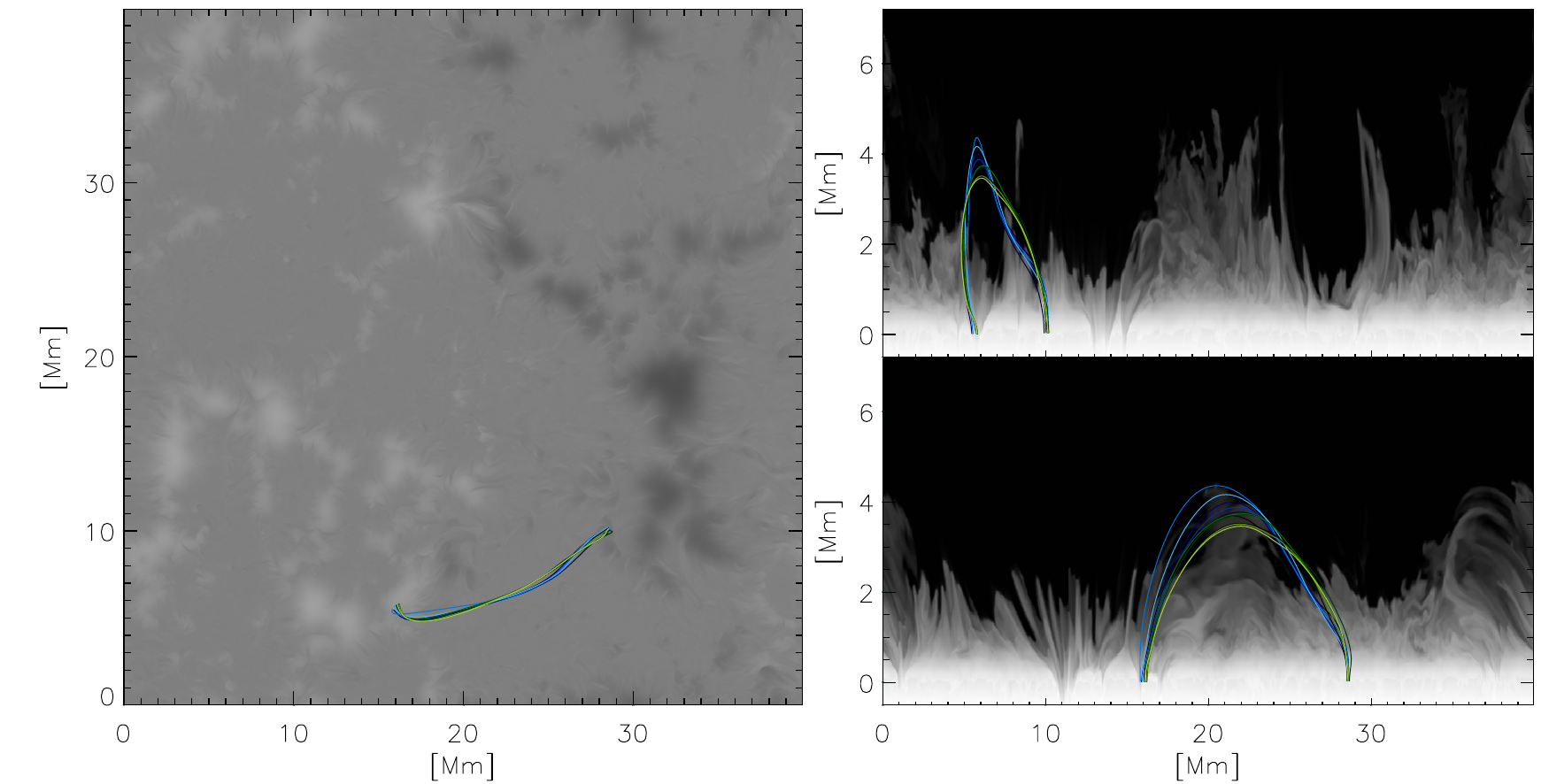}
\caption{Magnetic field lines traced for RREs F0 (blue lines) and F1 (green lines). Panel on the left shows top view with vertical photospheric magnetic field in the background. Panels on the right show projection of field lines in x (bottom) and y (top) planes with density cut at x,y coordinates of one of the initial seeds.}
\label{fig:F01_movie}
\end{figure*}
}
%\bibliography{rbes}{}
\bibliographystyle{aa.bst}

\end{document}